\documentclass[fleqn]{article}
\usepackage{espcrc2}
\usepackage{graphicx}
\usepackage{amsfonts}
\newcommand{\bel}{\begin{equation}}
\newcommand{\ee}{\end{equation}}
\def\rys#1#2#3{\begin{figure}[t]
      \vskip 3mm
      \centerline
      {
      \includegraphics*[width=0.5 \textwidth]{#1}
      }
      \caption{#2}
      \label{#3}
      \vskip 3mm
      \end{figure}
      }

\newcommand{\AmS}{{\protect\the\textfont2
  A\kern-.1667em\lower.5ex\hbox{M}\kern-.125emS}}
\title{WroNG -- Wroc\l aw Neutrino Generator of events for single pion production}
\author{Jan~T.~Sobczyk\address[IFT]{Institute of Theoretical Physics, Wroc\l aw University.\\
pl. M. Borna 9, 50-204 Wroc\l aw, Poland}\thanks{The authors were
supported by KBN grant 105/E-344/SPB/ICARUS/P-03/DZ211/2003-2005}
, Jaros\l aw A. Nowak\addressmark[IFT],
Krzysztof~M.~Graczyk\addressmark[IFT]}

\begin{document}

\begin{abstract}

We constructed a new Monte Carlo generator of events for neutrino
CC single pion production on free nucleon targets. The code uses
dynamical models of the DIS with the PDFs modified according to
the recent JLab data and of the $\Delta$ excitation. A comparison
with experimental data was done in three channels for the total
cross sections and for the distributions of events in invariant
hadronic mass.
\end{abstract}

\maketitle

\section{INTRODUCTION}

Neutrino-nucleon (or nucleus) CC interactions are described in a
framework of three different theoretical schemes: quasi-elastic,
resonance, and deeply inelastic (DIS). It is a nontrivial task to
put them together in a Monte Carlo generator of events. Most
problems arise in the resonance region. There are several models
of single pion production (SPP) due to resonance production but
some non-resonant background is also required. Then it is
necessary to join such a model with the DIS part which must be
extrapolated far away from the kinematical region in which it is
reliable. In existing MC codes \cite{Nuint} the Rein-Sehgal model
\cite{RS} is most often used to describe the dynamics of SPP. It
includes contributions from several resonances of the mass up to
$1.8~GeV$ added in the coherent way. The non-resonant background
is then added incoherently in order to fit the experimental data.
An important improvements to existing MC codes can come from
precise JLab experimental data from electron scattering
experiments \cite{Duality}. It is known how to construct good
experimental fits for the structure functions $F_1$ and $F_2$
which in the resonance kinematical region average over resonance
peaks \cite{Bodek}. In the neutrino-nucleon cross section the
axial structure function $F_3$ is also present whose modification
cannot be deduced from electron experiments. As a first guess one
can assume that the modification is analogous to that applied to
the $F_{1,2}$ and investigate consequences of such assumption.

\rys{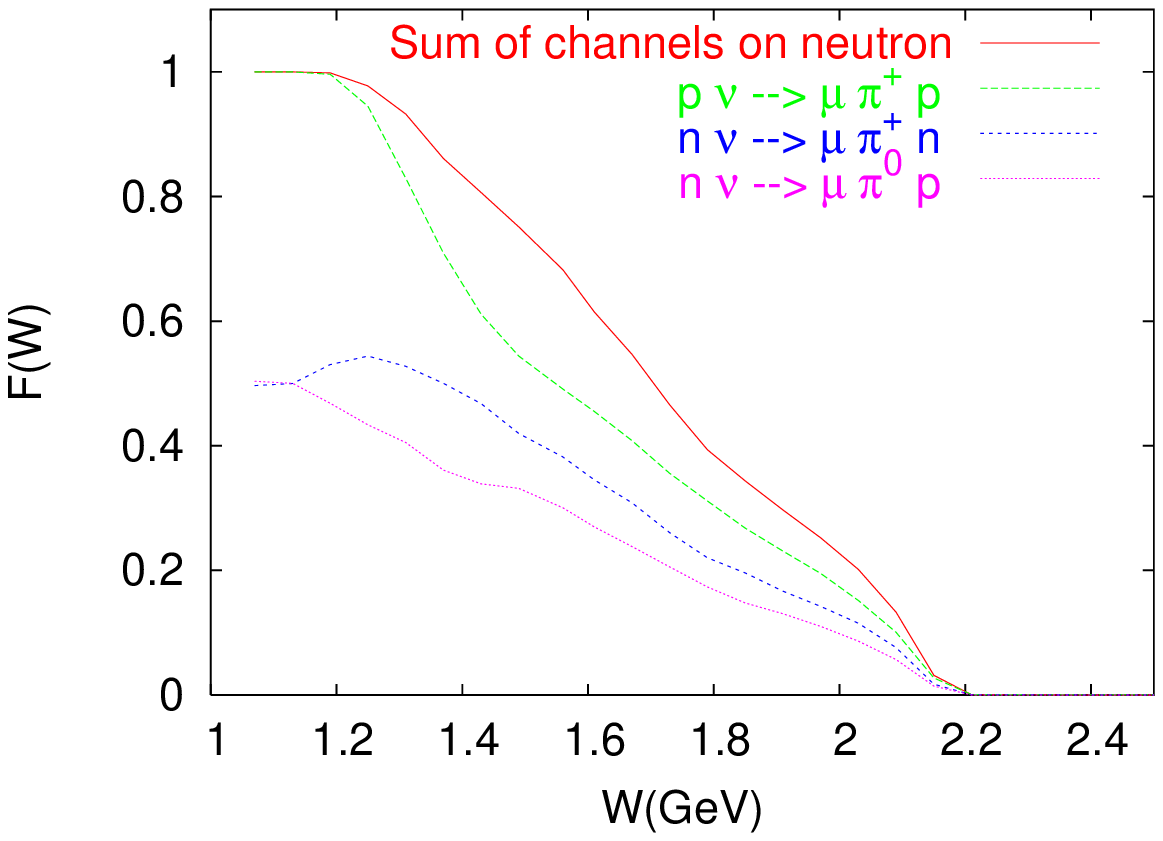}{Fraction of single pion production
contribution in overall Deep Inelastic Scattering cross
section}{spp_function}

\rys{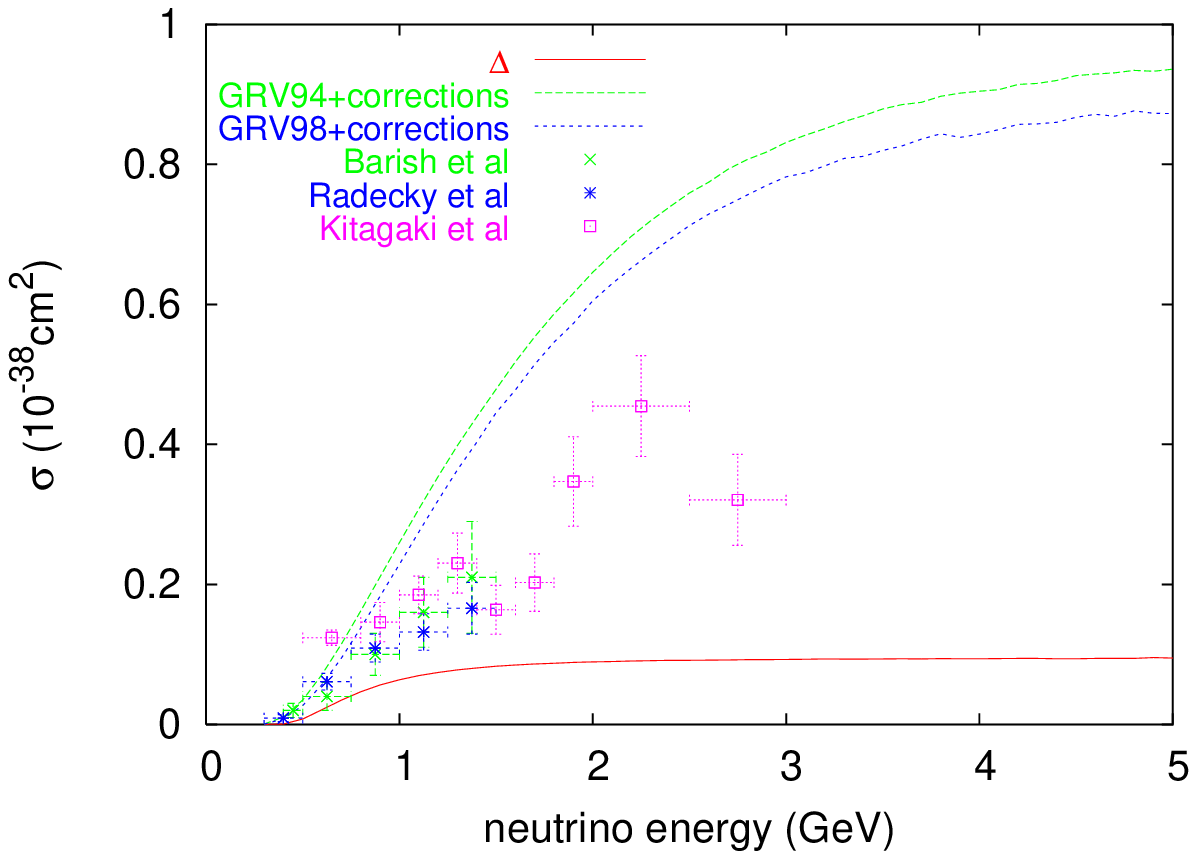}{Total cross section in the channel
$\nu_ \mu n \rightarrow \mu ^- \pi ^+ n$}{total_n_pionplus}

\rys{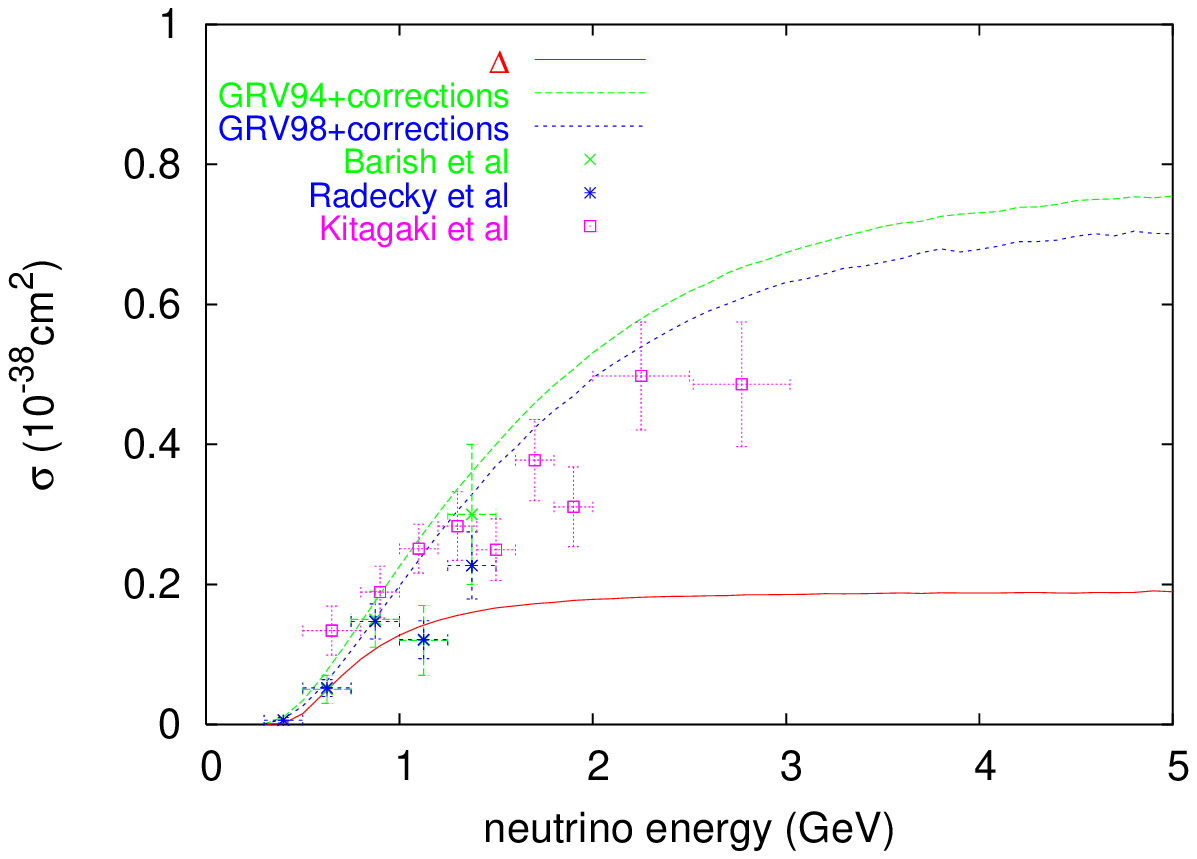}{Total cross section in the channel
$\nu_ \mu n \rightarrow \mu ^- \pi ^0 p$}{total_n_pionzero}

\rys{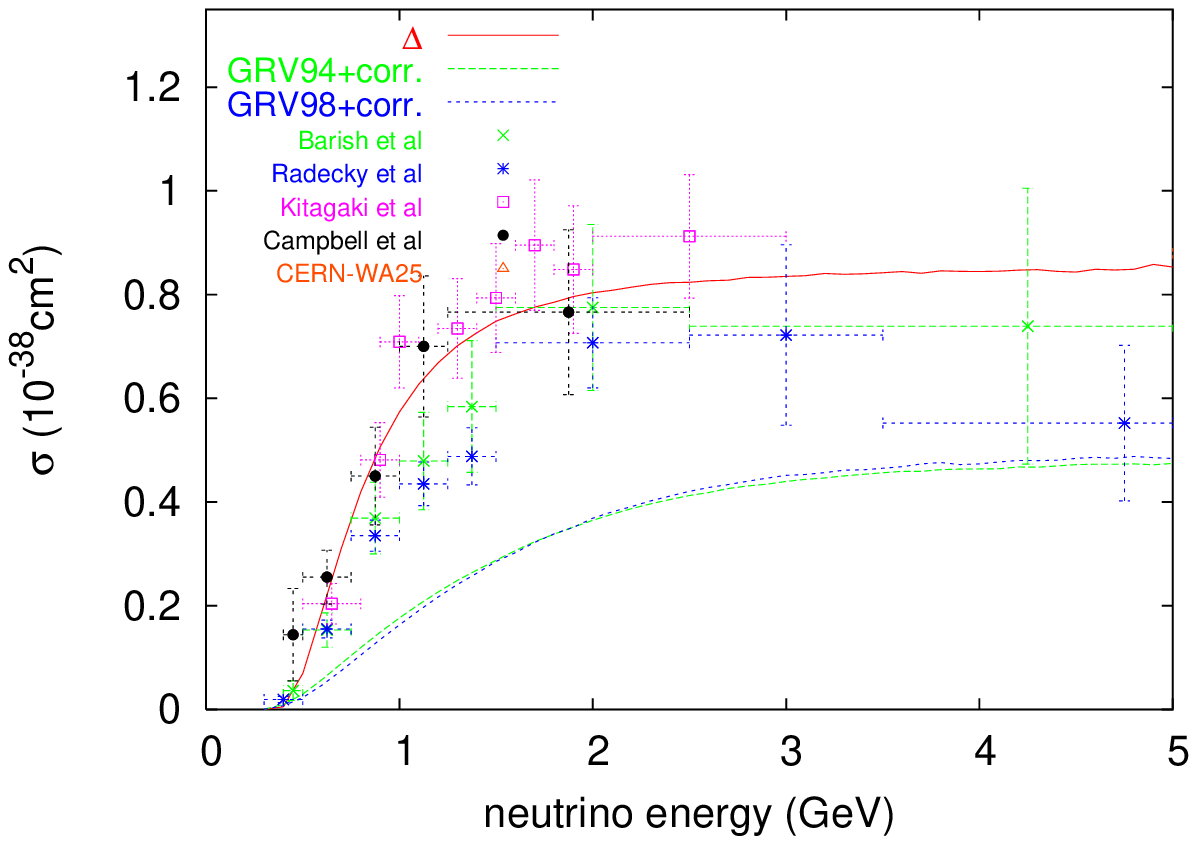}{Total cross section in the channel
$\nu_ \mu p \rightarrow \mu ^- \pi ^+ p$}{total_p_pionplus}

This was the starting point for our investigation. We shall
describe a construction of MC code WroNG (WROc\l aw Neutrino
Generator) for SPP which incorporates explicit $\Delta$ excitation
model and three exclusive SPP channels extracted from the DIS
formalism. We focused on SPP channels since inclusion of the
remaining dynamics (quasi-elastic channel and more inelastic
reactions described by DIS formalism) is straightforward. Our code
can supplement the NUX+FLUKA scheme which does not contain a
resonance contribution \cite{NuxFluka}.

In order to evaluate SPP in the framework of the DIS formalism we
introduced three functions (one for each channel) of kinematical
variables. The functions measure the probability that after
fragmentation and hadronisation the final hadronic state is that
of SPP. We obtained these functions from the NUX+FLUKA simulations
which are based on the LUND algorithm \cite{lund}. They turned out
to be monotonously decreasing functions of the hadronic mass $W$,
taking values in a range from $1$ to $0$ (see fig. 1).

The functions we introduced above have been used to define the
differential cross section for SPP in the DIS formalism (there are
three identical definitions for each reaction channel):

\begin{equation}\label{DISCrossSection}
{d^2\sigma^{DIS-SPP}  \over dWd\omega}= {d^2\sigma^{DIS} \over
dWd\omega}\cdot F (W)
\end{equation}

where $\sigma^{DIS}$ is the DIS inclusive cross section. $W$ and
$\omega$ denote hadronic mass and energy transfer respectively.

\clearpage
 In our work we had to solve two problems. The first was to
join $\Delta$ and DIS contributions. The second problem was to add
appropriate non-resonant contributions. In what follows by DIS we
mean exclusive SPP channels contained in inclusive DIS. We joined
two dynamical mechanisms in the cross section expression according
to the values of the hadronic invariant mass in the kinematically
allowed region. The basic idea was that for small (i.e. from the
threshold $W=M+m_{\pi}$ to about $1.4~GeV$) values of $W$ the
dynamics is that of $\Delta$ excitation while for larger $W$ the
dynamics is that of DIS. In order to make the transition smooth we
fixed a region from $W_1$ to $W_2$ in which the probability to
choose either of two dynamics changed linearly in the MC way.

\rys{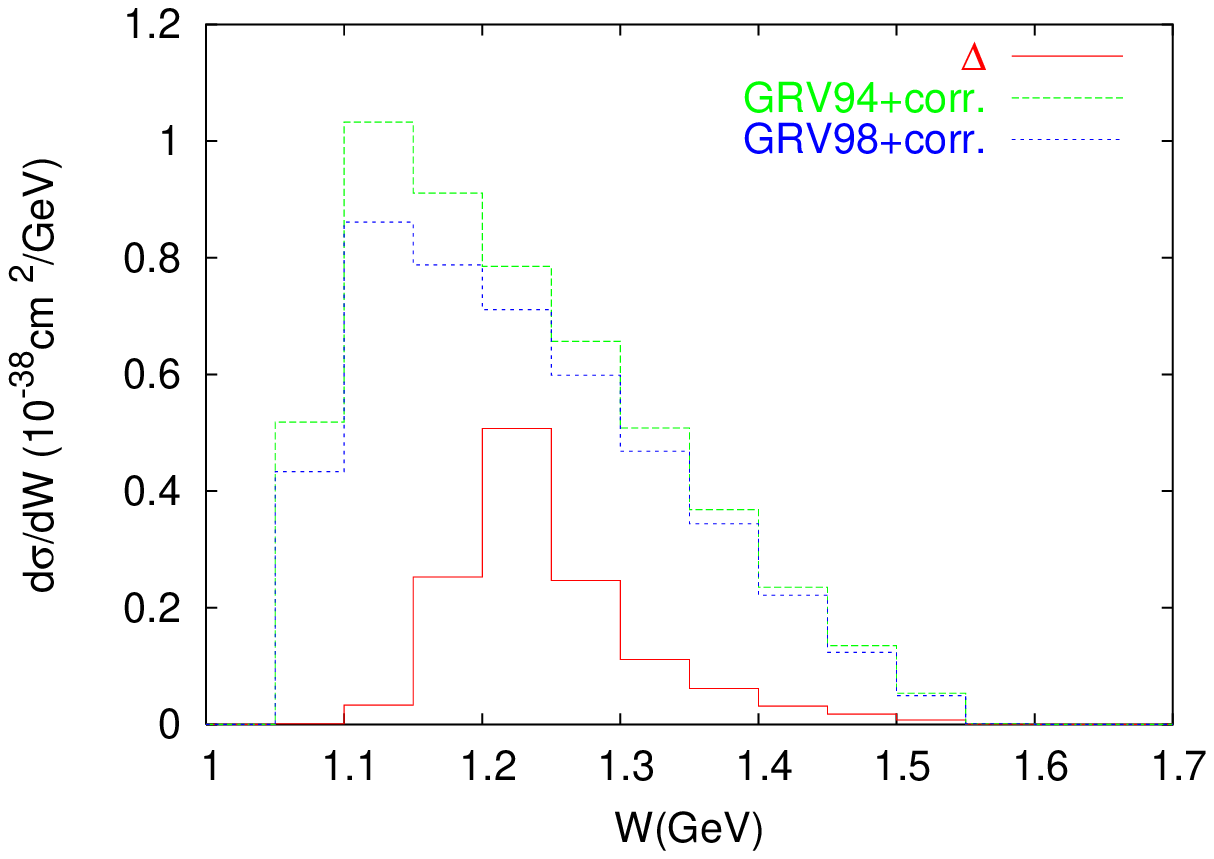}{Hadronic  mass distribution in the
channel $\nu_ \mu n \rightarrow \mu ^- \pi ^+ n$ at $E_\nu = 1
GeV$}{im_n_pionplus}

\rys{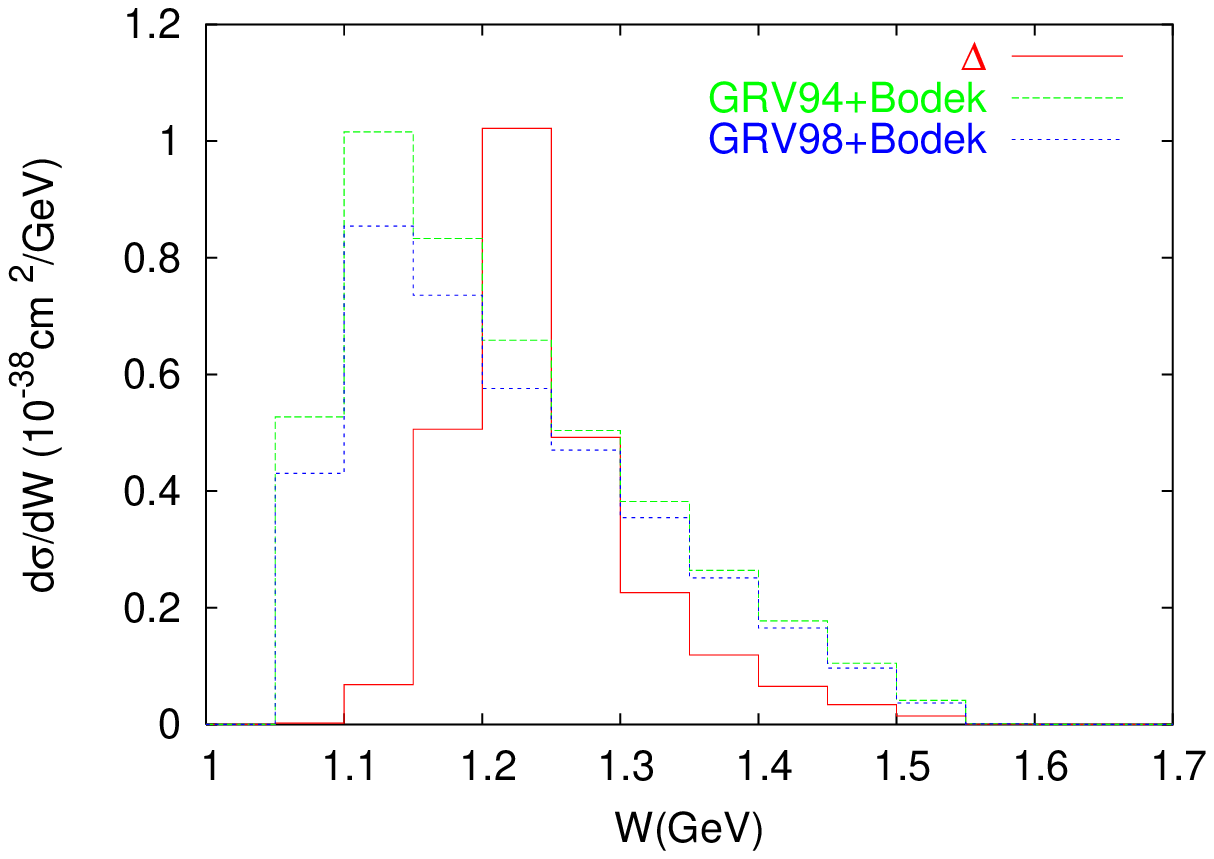}{Hadronic  mass distribution in the
channel $\nu_ \mu n \rightarrow \mu ^- \pi ^0 p$ at $E_\nu = 1
GeV$}{im_n_pionzero}

\rys{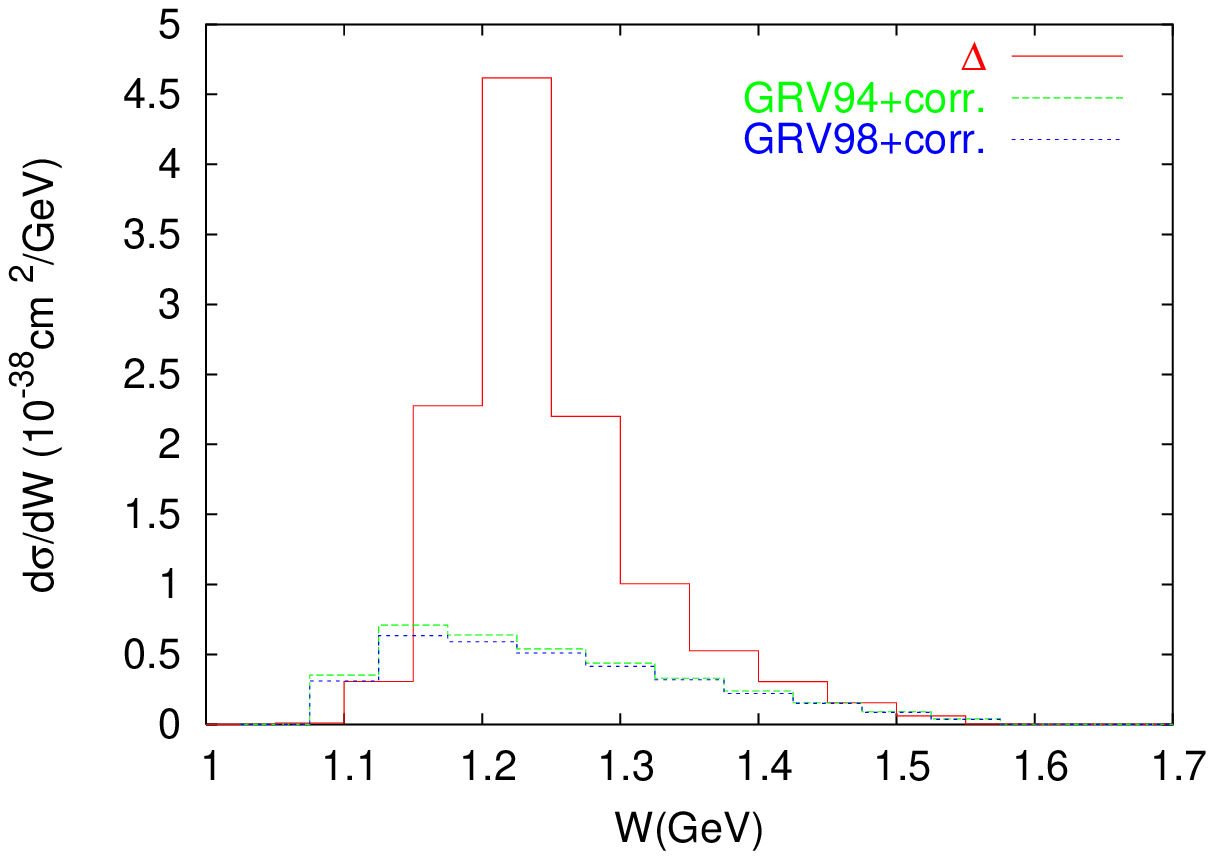}{Hadronic  mass distribution in the
channel $\nu_ \mu p \rightarrow \mu ^- \pi ^+ p$ at $E_\nu = 1
GeV$}{im_p_pionplus}

 We mimic the "non-resonant" background in the region
of small values of $W$ by an admixture of the DIS contribution. It
was done in the MC way and amount of the DIS contribution was
described by a parameter $\alpha$ a value of which was fixed by
making a comparison with experimental data separately in each
exclusive channel. To summarize the formula for the cross section
in each reaction channel can be written as:








\begin{equation}\label{CrossSection}
\begin{array}{l}
\frac{\displaystyle d^2 \sigma }{\displaystyle dWd\omega } = \theta \left( W_1  - W\right)\\
\\
\ \ \ \left( \alpha \frac{\displaystyle d^2 \sigma^{DIS-SPP}}
{\displaystyle dWd\omega } + \left( 1 - \alpha
\right)\frac{\displaystyle d^2 \sigma^{\Delta}} {\displaystyle
dWd\omega }\right)\\
\\
+ \theta \left( W - W_1  \right)\theta \left( W_2  - W \right)\\
\\
\ \ \ \left( \left( \alpha  + \left( 1 - \alpha
\right)\frac{\displaystyle W - W_1 }{\displaystyle W_2 -
W_1 } \right )\frac{\displaystyle d^2 \sigma^{DIS-SPP} }{\displaystyle dWd\omega } \right.\\
\\
\ \ \ \ \ +\left. \left( 1 - \alpha  \right)\frac{\displaystyle
W_2 - W}{\displaystyle W_2  - W_1 }
\frac{\displaystyle d^2 \sigma^\Delta}{\displaystyle dWd\omega } \right)\\

\\
+\theta \left( W - W_2  \right)\frac{\displaystyle d^2 \sigma
^{DIS-SPP} }{\displaystyle dWd\omega }
\end{array}
\end{equation}

\section{RESULTS}

\rys{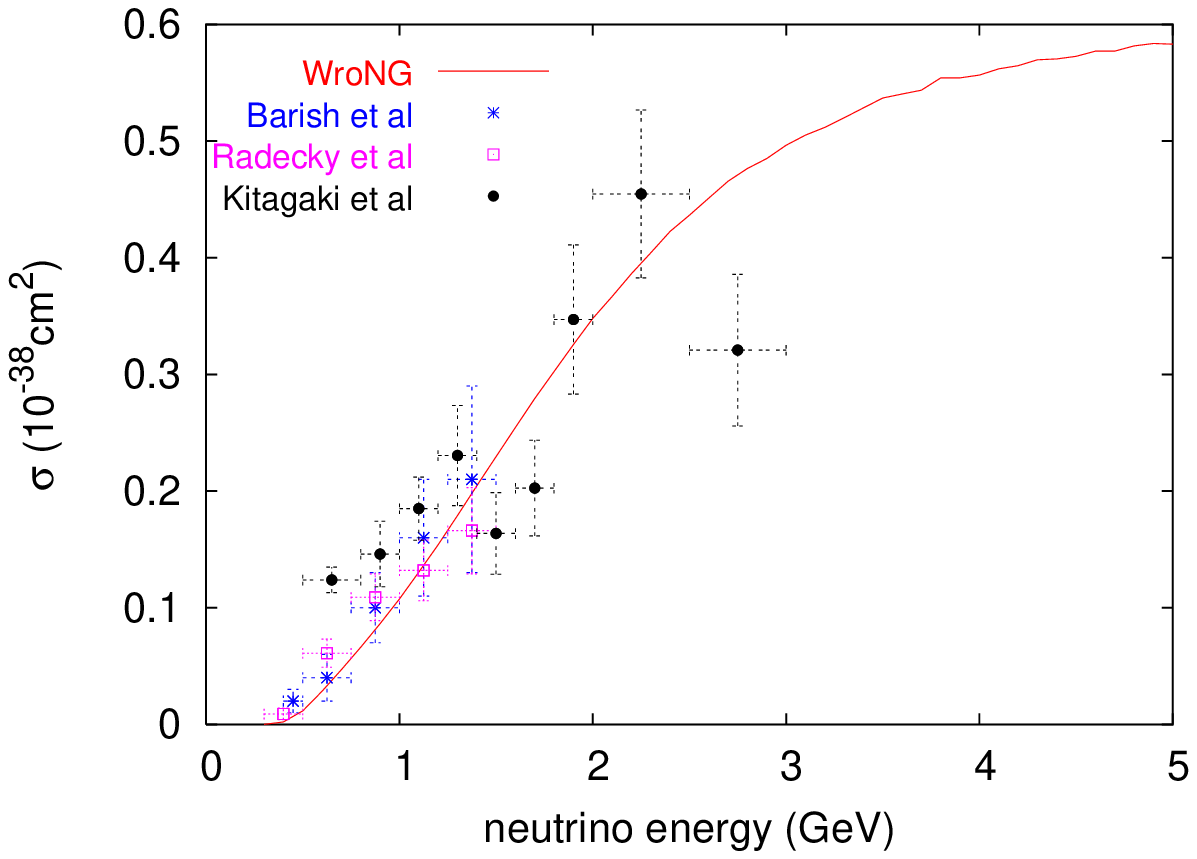}{WroNG predictions in the channel
$\nu_ \mu n \rightarrow \mu ^- \pi ^+ n$ and experimental data
(total cross section) from \cite{BRData}
\cite{Kitagaki}}{final_cross_n_pionplus}

\rys{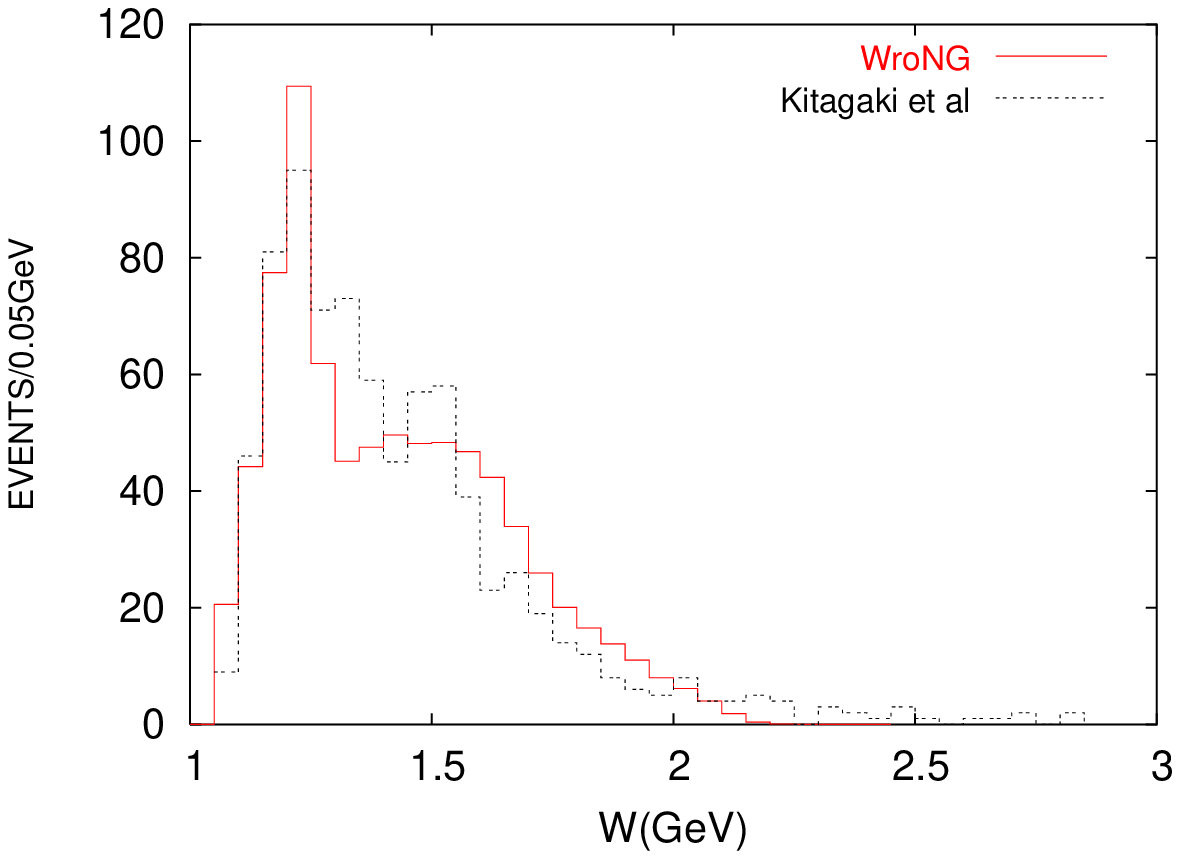}{WroNG predictions in the
channel$\nu_ \mu n \rightarrow \mu ^- \pi ^+ n$ and experimental
data (hadronic mass distribution) from
\cite{Kitagaki}}{final_diff_n_pionplus}

\rys{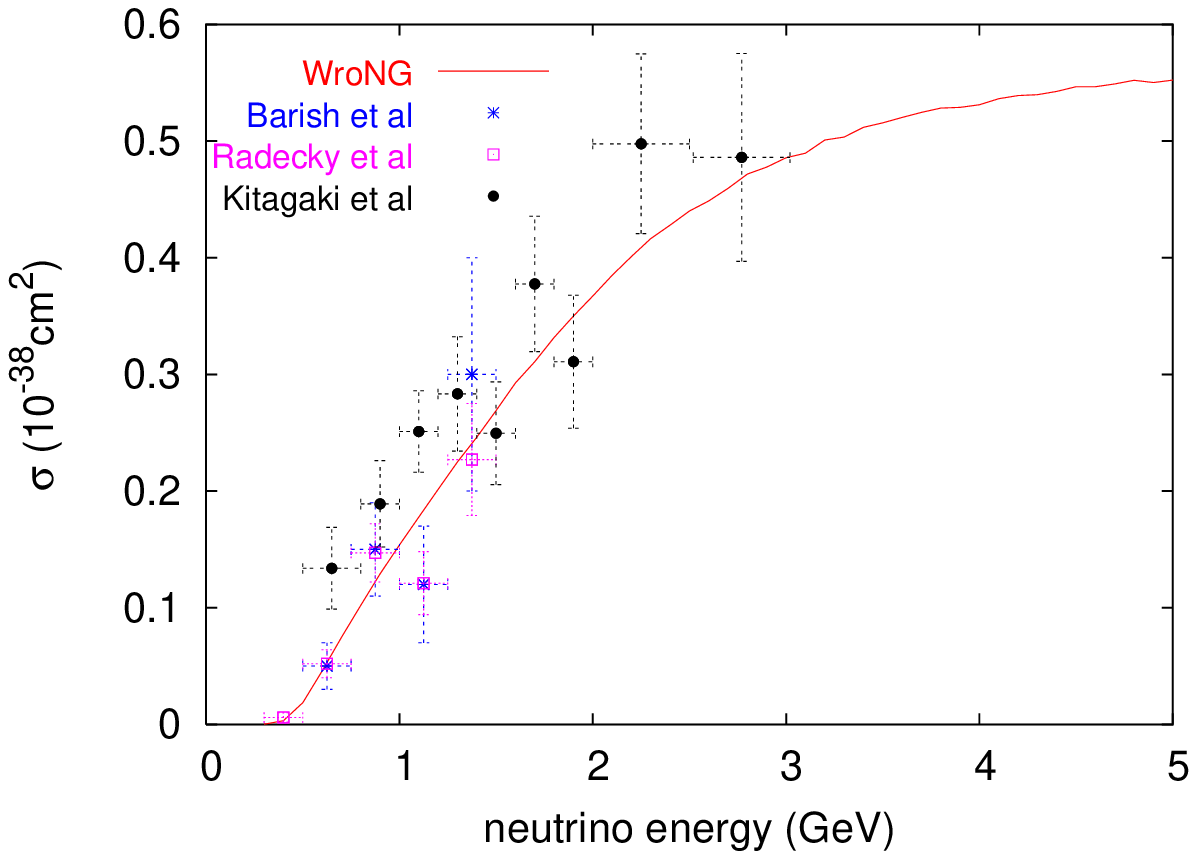}{WroNG predictions in the channel
$\nu_ \mu n \rightarrow \mu ^- \pi ^0 p$ and experimental data
(total cross section) from \cite{BRData}
\cite{Kitagaki}}{final_cross_n_pionzero}

\rys{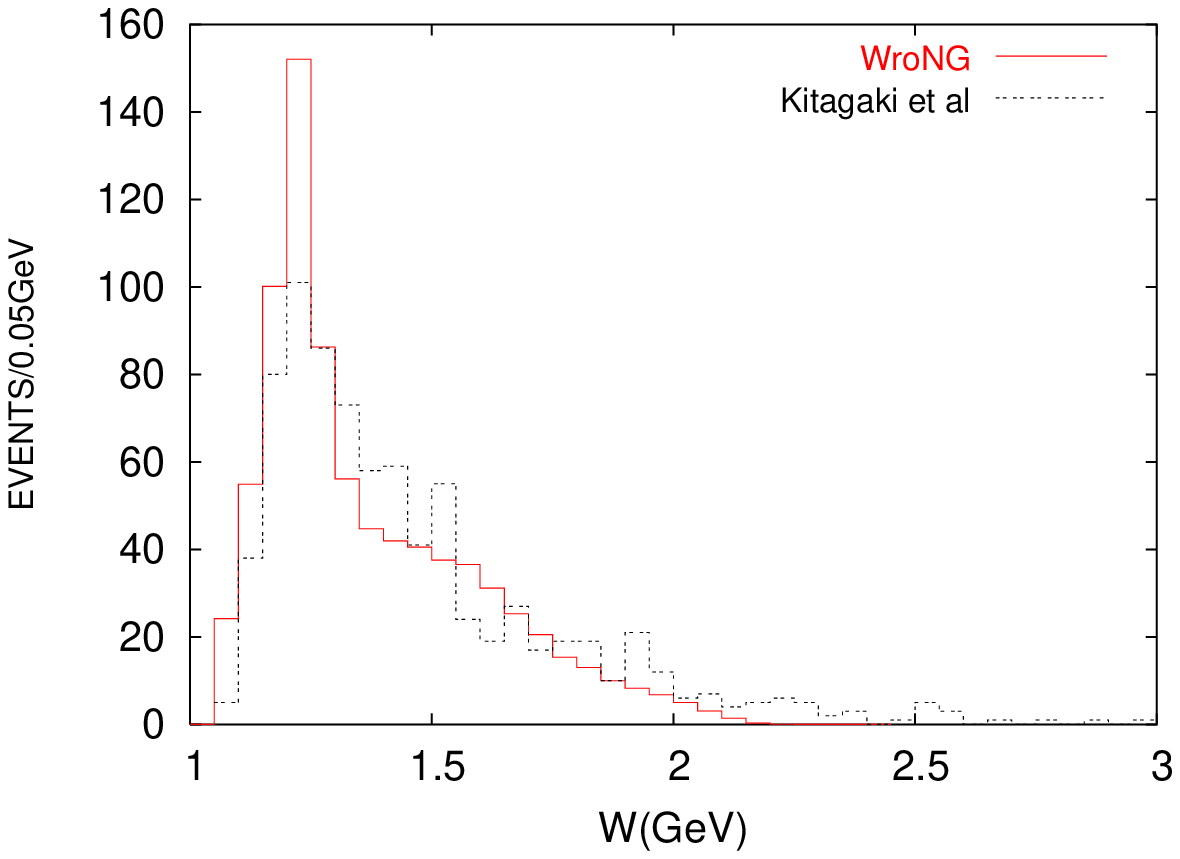}{WroNG predictions in the channel
$\nu_ \mu n \rightarrow \mu ^- \pi ^0 p$ and experimental data
(hadronic mass distribution) from
\cite{Kitagaki}}{final_diff_n_pionzero}

\rys{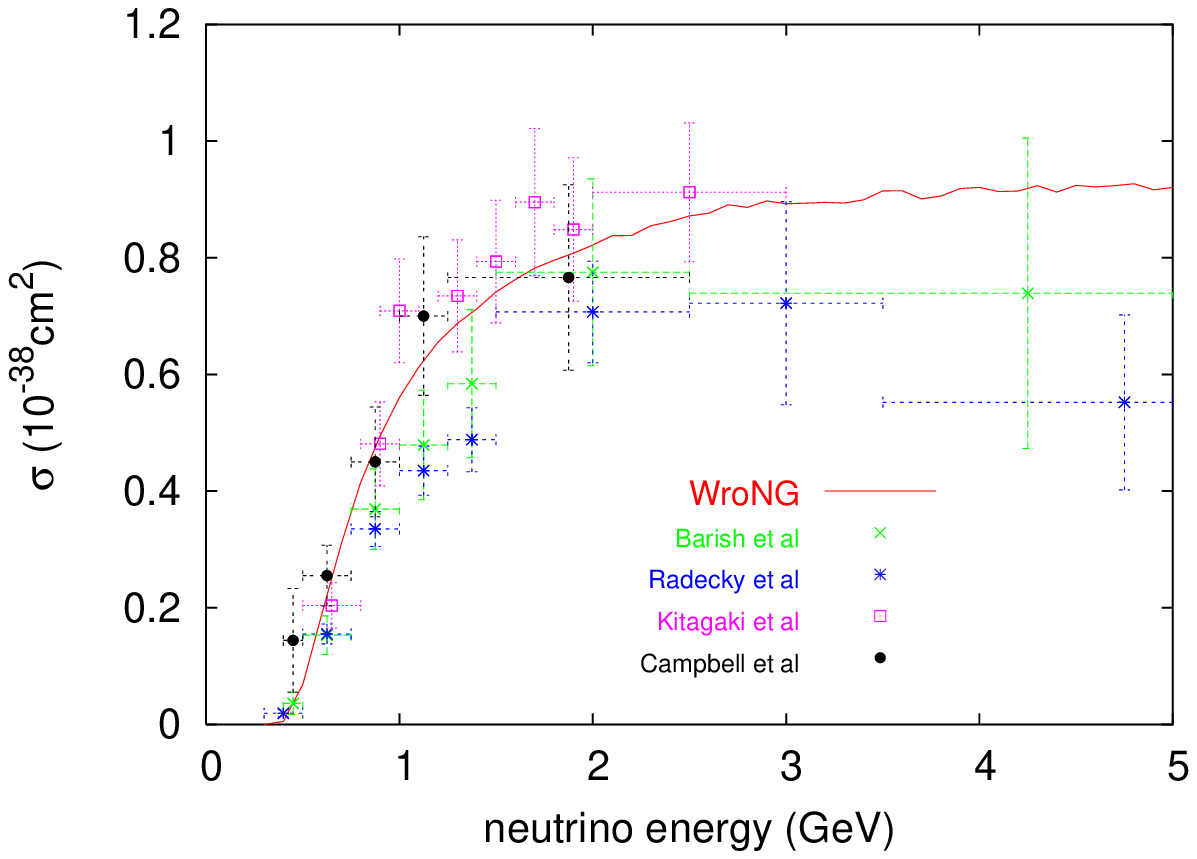}{WroNG predictions in the channel
$\nu_ \mu p \rightarrow \mu ^- \pi ^+ p$ and experimental data
(total cross section) from \cite{BRData}
\cite{Kitagaki}}{final_cross_p_pionplus}

\rys{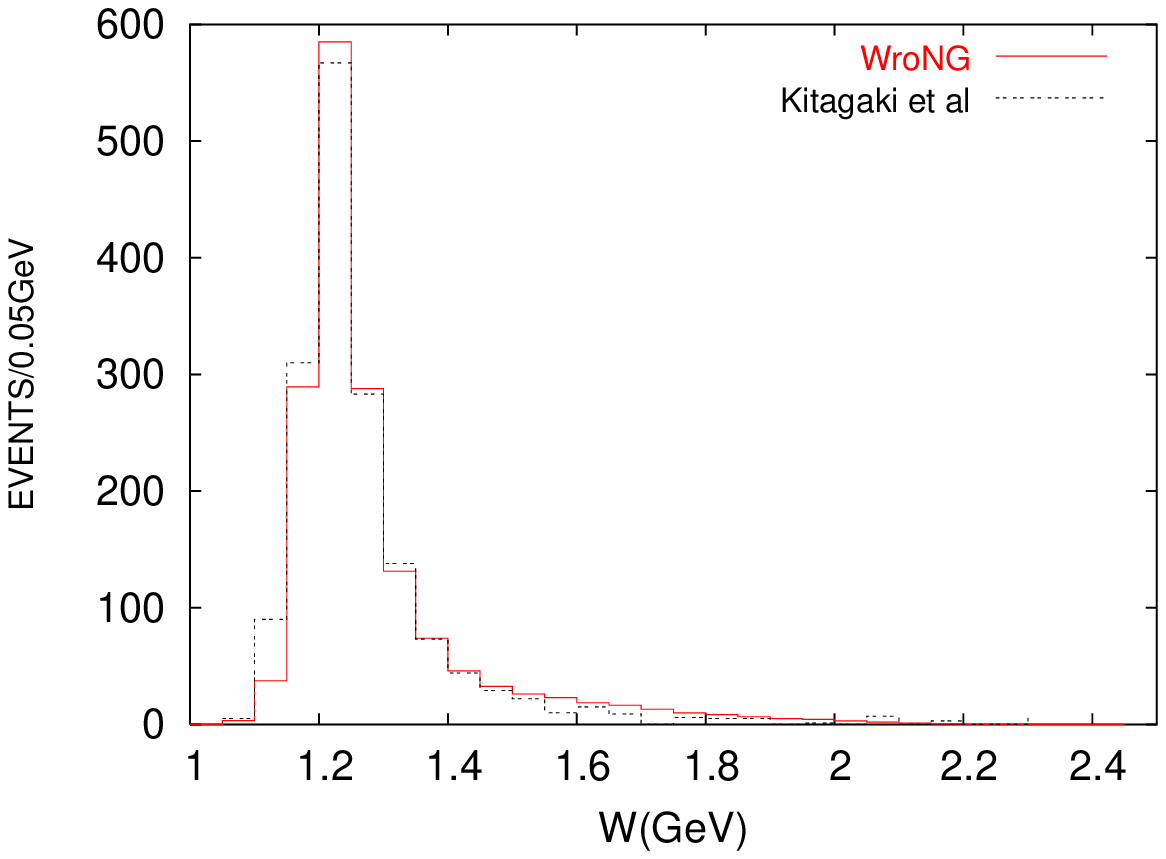}{WroNG predictions in the channel
$\nu_ \mu p \rightarrow \mu ^- \pi ^+ p$ and experimental data
(hadronic mass distribution) from
\cite{Kitagaki}}{final_diff_p_pionplus}

First we present basic ingredients of our construction: the
$\Delta$ excitation model (taken from the Marteau model
\cite{Marteau}) and the model of SPP based on DIS. We show the
total cross sections in three reaction channels (fig.~2-4) with
experimental points taken from papers: \cite{BRData},
\cite{Kitagaki}. We also show distributions of events in the
hadronic mass for neutrinos of energy $1~GeV$ (fig.~5-7). In the
DIS part we did computations for two sets of PDF: modifications of
GRV94 and GRV98. Differences between them are very small and in
what follows we restrict ourselves to GRV98 with corrections.

In the channel $\nu_{\mu} n\rightarrow \mu^- \pi^+ n$ the DIS
cross section is much bigger than for the $\Delta$ excitation one
(see fig.~\ref{total_n_pionplus}). If compared with experimental
data we see that the DIS predictions are above experimentally
observed while those of the $\Delta$ model below. In
fig.~\ref{im_n_pionplus} the DIS differential cross section of
hadronic mass is bigger than of the $\Delta$ in the whole
kinematical domain.

In the channel $\nu_{\mu} n\rightarrow \mu^- \pi^0 p$ the
situation is quite different (see fig.~\ref{total_n_pionzero}). At
smaller values of the neutrino energy $E$ (about $1~GeV$) both
models predict similar values of the total cross section close to
experimentally measured. For higher values of neutrino energy the
DIS predictions agree with the experimental data while the
$\Delta$ excitation model predictions are too low. In
fig.~\ref{im_n_pionzero} differential cross sections reveal that
while the cross sections are similar in the total area below the
curves, the shapes of the hadronic mass distributions of events
are very different. In this channel we see a nice manifestation
of the quark-hadron duality: the DIS contribution average over
resonance peak.

In the channel $\nu_{\mu} p\rightarrow \mu^- \pi^+ p$ the $\Delta$
excitation model predicts much higher values of the cross section
which are also close to the experimentally measured (see
fig.~\ref{total_p_pionplus}, \ref{im_p_pionplus}).

We conclude that each channel has its unique features and has to
be treated independently. We also note that quark-hadron duality
is seen in only one SPP channel. It suggests that modifications of
PDF we applied in our computations are not yet good enough.

Many choices for $W_1, W_2$ for each channel separately were
checked in order to find the most suitable one. The results did
not depend much on the choice and we fixed for all three channels:
$W_1=1.3~GeV$ and $W_2=1.6~GeV$.

Total cross sections depend in a substantial way on $\alpha$. In
the channel $\nu_{\mu} p\rightarrow \mu^- \pi^+ p$ an increase of
$\alpha$ makes the cross section smaller, in other two channels
the dependence is opposite. We also looked at distributions of
events in the hadronic mass. We compared our MC results with
Kitagaki et al. data \cite{Kitagaki} because they have the best
statistics. We took into account experimentally reconstructed
spectra of neutrinos and we produced samples of events with the
same spectrum. The number of events has been chosen to be the
same as in the oryginal experiment so that this part of the
analysis applied only to the shapes of hadronic mass distributions
of events.

In two channels on the neutron the value of $\alpha$ determines
the height of the $\Delta$ resonance peak. The transition region
$(W_1,W_2)$ can manifest itself as a resonance-like peak at values
of $W$ close to $W_2$. It is only by chance that such higher
resonance peak can be seen in the experimental data. With an
increase of $\alpha$ the resonance peak becomes lower but at the
expense of too many events at lower values of hadronic mass and in
clear contradiction with experimental data. The best value of
$\alpha$ is a compromise between two described tendencies.

In figs.~8-13 we show a comparison of our best qualitative fits
with Kitagaki et al. experimental data in three reaction channels.
We took $\alpha =0.2$ for $\nu_{\mu} n\rightarrow \mu^- \pi^+ n$
channel, $\alpha =0.3$ for $\nu_{\mu} n\rightarrow \mu^- \pi^0 p$
channel and $\alpha =0$ for $\nu_{\mu} p\rightarrow \mu^- \pi^+ p$
channel. Taking into account the simplicity of our construction we
find the agreement satisfactory.

We plan to improve WroNG in the following way:
\begin{itemize}
\item a non-resonant background will be introduced in
theoretically justified way using Fogli-Nardulli
\cite{FogliNardulli} or Lee-Sato \cite{LeeSato} results

\item three functions $F_{SPP}(W)$ will be derived directly from
PYTHIA/JETSET or LEPTO

\item the model will be extended to describe SPP on nucleus
targets
\end{itemize}

\end{document}